\documentclass[aps,prd,twocolumn,superscriptaddress,nofootinbib,10pt]{revtex4-2}
\usepackage[paperwidth=21cm,paperheight=29.7cm,top=2.54cm,bottom=2.54cm,left=2cm,right=2cm]{geometry}
\usepackage[colorlinks,linkcolor=blue,anchorcolor=blue,citecolor=blue,urlcolor=blue]{hyperref}
\usepackage{graphicx,subfigure}%dcolumn}
\usepackage[figuresright]{rotating}
\usepackage{amsmath,amsfonts,amssymb,bm}
\usepackage{array,enumitem,multirow}%,makecell}
\usepackage{acronym}
\usepackage{makecell}
\newcommand{\be}{\begin{equation}}
\newcommand{\ee}{\end{equation}}
\newcommand{\bea}{\begin{eqnarray}}
\newcommand{\eea}{\end{eqnarray}}
\newcommand{\TQC}{MOE Key Laboratory of TianQin Mission, TianQin Research Center for Gravitational Physics \&  School of Physics and Astronomy, Frontiers Science Center for TianQin, Gravitational Wave Research Center of CNSA, Sun Yat-sen University (Zhuhai Campus), Zhuhai 519082, China.}
\newacro{GR}{general relativity}
\newacro{GW}{gravitational wave}
\newacro{MG}{modified gravity theory}
\newacro{BH}{Black hole}
\newacro{PN}{post-Newtonion}
\newacro{ppE}{parameterized post-Einsteinian}
\newacro{GCB}{galactic ultra-compact binary}
\newacro{SBHB}{stellar-mass black hole binary}
\newacro{MBHB}{massive black hole binary}
\newacro{BHB}{black hole binary}
\newacro{IMBHB}{intermediate-mass black hole binary}
\newacro{EMRI}{extreme mass ratio inspiral}
\newacro{IMRI}{intermediate mass ratio inspiral}
\newacro{SGWB}{stochastic gravitational wave background}
\newacro{MECO}{minimal energy circular orbit}
\newacro{FAR}{false alarm rate}
\newacro{CE}{Cosmic Explorer}
\newacro{ET}{Einstein Telescope}
\newacro{LISA}{Laser Interferometer Space Antenna}
\newacro{EdGB}{Einstein-dilaton Gauss-Bonnet}
\newacro{dCS}{dynamic Chern-Simons}
\newacro{SNR}{signal-to-noise ratio}
\newacro{FIM}{Fisher Information Matrix}
\newacro{ISCO}{innermost stable circular orbit}
\newacro{NSBH}{neutron star-black hole binary}
\newacro{MCMC}{Markov Chain Monte Carlo}
\newacro{QNM}{quasi-nomral mode}
\newacro{LVK}{the LIGO-Virgo-Kagra collaboration}
\newacro{SWSH}{spin-weighted spheroidal harmonics}

\begin{document}

\preprint{APS/123-QED}

\title{On the detectability and resolvability of quasi-normal modes with space-based gravitational wave detectors}

\author{Changfu Shi}
\email{Email: shichf6@mail.sysu.edu.cn (Corresponding author)}
\author{Qingfei Zhang}
\author{Jianwei Mei}
\affiliation{\TQC}

\date{\today}% It is always \today, today,

\begin{abstract}
The detection of quasi-normal modes during the ringdown phase is a crucial method for testing the no-hair theorem. In this paper, the detectability and resolvability of multiple quasi-normal modes using space-based gravitational wave detectors have been analyzed. The results indicate that TianQin and LISA have the potential to detect and resolve a series of modes, including six fundamental modes, one overtone, and two nonlinear second-order modes. Furthermore, the analysis of systematic errors in the waveform suggests that even modes such as (3,3,1) and (4,3,0), which are unlikely to be directly detected and resolved, need to be taken into account in the ringdown waveform templates.
\end{abstract}

\maketitle

%\tableofcontents
\section{Introduction\label{int}}
The direct detection of \acp{GW} has not only opened a new window into the universe \cite{LIGOScientific:2016aoc} but also provided novel methods for testing the nature of gravity and black holes in highly dynamic and nonlinear regions. Since the first detection of the GW event GW150914 \cite{LIGOScientific:2016aoc,LIGOScientific:2021djp}, \ac{LVK} has announced a total of 90 GW events originating from compact binary mergers \cite{LIGOScientific:2018mvr,LIGOScientific:2020ibl,LIGOScientific:2021djp}. These discoveries have sparked extensive discussions across astronomy \cite{Riles:2017dki,Greco:2018gba,LIGOScientific:2019gag}, cosmology \cite{LIGOScientific:2017adf,Hotokezaka:2018dfi,LIGOScientific:2018gmd}, and fundamental physics \cite{LIGOScientific:2016lio,Abbott:2018lct,LIGOScientific:2019fpa,LIGOScientific:2020tif,LIGOScientific:2021sio,Perkins:2021mhb,Wang:2021jfc,Niu:2021nic,Wang:2021ctl,Kobakhidze:2016cqh,Yunes:2016jcc}, with one of the most notable applications being the tests of the no-hair theorem.

The detection of the ringdown phase presents an experimental avenue for testing the no-hair theorem \cite{Dreyer:2003bv}. Following the coalescence of a binary system, the remnant object rapidly transitions from a highly perturbed state to a Kerr black hole \cite{Leaver:1986gd}, emitting a series of damped oscillating signals known as the \acp{QNM} of the remnant\cite{Berti2009}. Each mode is uniquely characterized by its oscillation frequency and decay time, parameters that are solely determined by the total mass and spin of the remnant black hole \cite{Teukolsky:1973ha}. The no-hair theorem can be rigorously tested by measuring multiple QNMs and analyzing the consistency between the mass and spin inferred from each individual mode \cite{Echeverria:1989hg,Berti:2018vdi}. The crux of this test lies in the successful detection and resolution of multiple QNMs.

However, the current sensitivity of ground-based GW detectors limits the detection of QNMs beyond the dominant mode, resulting in relatively low \acp{SNR} and ongoing debates about the definitive evidence for these modes in the data \cite{Isi:2019aib,CalderonBustillo:2020rmh,Finch:2022ynt,Ma:2023vvr,Crisostomi:2023tle,Carullo:2019flw,Isi:2021iql,Isi:2022mhy,Isi:2023nif,Carullo:2023gtf,Wang:2023xsy,Capano:2021etf,Capano:2022zqm,Cotesta:2022pci,Siegel:2023lxl,Wang:2023mst,Wang:2021hfd,Correia:2023bfn}. With technological advancements, the upcoming third-generation of ground-based and space-based GW detectors are expected to possess the capability to detect a broader range of GW signals with significantly higher SNRs \cite{Punturo:2010zz,Reitze:2019iox,Wang:2019ryf,Fan:2020zhy,Huang:2020rjf,Liu:2020eko,Liang:2021bde,TianQin:2020hid,Torres-Orjuela:2023hfd,Maggiore:2019uih,Babak:2017tow,LISA:2022yao,LISA:2022kgy,LISACosmologyWorkingGroup:2022jok}. Those progress will make the detection of higher-order \acp{QNM} a realistic possibility.

While most previous work on this topic has focused on the detection capabilities of a few fundamental modes. Berti et al. provided the first quantitative assessment of the detection capability of space-based GW detectors for the ringdown QNMs \cite{Berti2006a}. They analyzed the capability of LISA to resolve three subleading QNMs from dominant mode as a function of the mass ratio. By fitting this result using polynomials, Berti et al \cite{Berti:2016lat}. further analyzed the number of events for which the third generation of ground-based detectors and space-based GW detectors could resolve at least two modes, utilizing black hole astronomical models. Gossan et al. \cite{Gossan:2011ha}, building on the fitting functions relating the amplitudes of several fundamental QNMs introduced by Kamaretsos et al. \cite{Kamaretsos:2011um}, investigated the capability of eLISA \cite{Amaro-Seoane:2012aqc} and the Einstein Telescope \cite{Punturo:2010zz} to test the no-hair theorem as a function of the luminosity distance. Shi et al. \cite{Shi2019} conducted a more detailed analysis with the help of astronomical models of massive black holes for TianQin \cite{Luo:2015ght} and LISA \cite{LISA:2017pwj}. Besides, only Baibhav et al. \cite{Baibhav:2018rfk} has addressed the capacities of future detector to detect higher-order fundamental modes.

However, while space-based GW detectors are anticipated to have the capability to directly detect overtones \cite{Giesler2019,Ota:2019bzl} and second-order modes \cite{Mitman:2022qdl,Cheung:2022rbm,Ioka2007}, quantitative analyses for these mode are still lacking. Currently, the analysis of mode resolvability is limited to scenarios where only the dominant mode and one additional mode are present in the signal. The expectation is that space-based GW detectors will be able to simultaneously detect a series of \acp{QNM}, necessitating further expansion of the analysis of this issue. 

The main target of this paper is to analyze the detectabilities and resolvabilities of multiple QNMs with space-based GW detectors, using TianQin and LISA as specific examples. Using the results of amplitude fitting from \cite{London2014} and \cite{Cheung:2022rbm}, we simulate the ringdown signal as a linear superposition of 11 QNMs, encompassing seven fundamental modes, two overtones, and two nonlinear second-order modes. By calculating the SNRs and the parameter estimation accuracy for each individual mode, we investigate the parametric dependence of the detectabilities and resolvabilities of multiple QNMs on source parameters. Additionally, utilizing three astrophysical population models of massive black holes, we estimate the expected detection and resolved numbers. The results indicate that TianQin and LISA exhibit considerable potential in detecting and resolving the majority of the modes considered in this paper, including one overtone and two nonlinear second-order modes.

The paper is structured as follows: In Section \ref{sec:me}, we introduce all the necessary components required for the calculations, including the waveform for the ringdown signals \ref{sec:rd} and the statistical method \ref{sec:sta}. In Section \ref{sec:result}, we present the main results, encompassing detectability \ref{sec:de}, resolvability \ref{sec:resolve}, and waveform modeling requirement \ref{sec:modeling}. A concise conclusion is provided in Section \ref{sec:conclusion}. Throughout this paper, we adopt the convention where $G=c=1$.

\section{Method\label{sec:me}}

This section outlines the construction of the ringdown signal and the statistical methods employed in this study.

\subsection{\label{sec:rd} Ringdown Signal}

The evolution of a black hole binary system can be broadly categorized into three phases: inspiral, merger, and ringdown. The ringdown phase occurs subsequent to the merger, where the newly formed compact object is in a highly perturbed state. The waveforms during this phase can be effectively modeled as a superposition of \acp{QNM} that oscillate and decay over time. These QNMs, whose oscillation frequencies and decay times are determined by solving the linear perturbation equations with suitable boundary conditions. Assuming the remnant object to be a Kerr black hole, the linear QNMs are typically denoted by three indices, \( (l, m, n) \), where \( l \) represents the harmonic index with values \( l = 2, 3, 4, \dots \), \( m \) denotes the azimuthal number with possible values \( m = 0, \pm 1, \pm 2, \dots \), and \( n \) is the overtone number, \( n = 0, 1, 2, \dots \). The fundamental modes, characterized by \( n = 0 \), are generally more detectable and exhibit longer damping times compared to the higher overtones with \( n \geq 1 \) .

The ringdown waveform considered in this paper is approximated by the following:
\begin{equation}
	\label{eq:waveform}
	h_{\text{Rd}} = \frac{M_z}{D_L} \sum_{l,m,n} \ _{-2}S_{l,m,n} A_{l,m,n} e^{i \widetilde{\omega}_{l,m,n} (t - t_0)}
\end{equation}
Here, \( M_z \) and \( D_L \) represent the red-shifted mass of the remnant and the luminosity distance of the system, respectively. \( A_{l,m,n} \), \( _{-2}S_{l,m,n} \), and \( \widetilde{\omega}_{l,m,n} = \omega_{l,m,n} + i / \tau_{l,m,n} \) are the amplitude, the -2 weighted spin spheroidal harmonics, and the complex frequency of the corresponding QNM, where \( \omega_{l,m,n} \) and \( \tau_{l,m,n} \) are the oscillation frequency and damping time, respectively.

The excitation amplitudes of the QNMs, \( A_{l,m,n} \), are dependent on the parameters of the progenitor binaries. Currently, these cannot be calculated analytically. Utilizing numerical relativity simulations, several efforts have been made to fit these amplitude functions to numerical results. In this paper, we employ the fitted formulas for linear mode amplitudes proposed by London et al. \cite{London2014}, which are applicable for systems with non-spinning progenitor black holes. Due to limitations imposed by the fitting results, specifically Eq. (1) in \cite{London2014}, we will consider only 9 linear QNMs in this paper, corresponding to the set \( \{l, m, n\} \in \{\{2, 2, 0\}, \{2, 2, 1\}, \{2, 1, 0\}, \{3, 3, 0\}, \{3, 3, 1\}, \{3, 2, 0\}, \\\{4, 4, 0\}, \{4, 3, 0\}, \{5, 5, 0\} \} \).

The -2 spin-weighted spheroidal harmonics \( _{-2}S_{l,m,n}(\iota, \phi, j M_z \widetilde{\omega}) \) arise from separating variables in the partial differential equations that describe the propagation of a spin-2 field in a Kerr background \cite{Leaver1986}. \( \iota \) is the inclination angle, \( \phi \) is the azimuthal angle of the source, and \( j \) is the dimensionless spin parameter. The spin-weighted spheroidal harmonics can be expanded in the basis of spherical harmonics \cite{Berti2004}, as follows:
\begin{align}
	_{-2}S_{l,m,n}(\theta, \phi, j M_z \widetilde{\omega})
	= \sum_{l' = l_{\text{min}}}^{l_{\text{max}}}
	C_{l'lm}(j M_z \widetilde{\omega}) _{-2}Y_{l' m}(\theta, \phi),
\end{align}
where \( C_{l'lm}(a \omega) \) is the spherical-spheroidal mixing coefficient, which is challenging to calculate analytically. In this paper, we use Stein's Python package, \textit{qnm} \cite{Stein:2019mop}, to obtain the mixing coefficients numerically.

The complex frequencies \( \widetilde{\omega}_{l,m,n} \), along with the oscillation frequencies \( \omega_{l,m,n} \) and damping times \( \tau_{l,m,n} \), are the eigenmodes of the radial part of the Teukolsky equation and depend solely on the mass \( M_z \) and spin \( j \) of the final Kerr black hole. These are quite difficult to compute analytically, and we will use the fitting formula proposed in \cite{Berti2006a} for our calculations.

We also consider the second-order QNMs, also known as quadratic QNMs. The contribution of the second-order modes to the waveform follows the same form as Eq. \eqref{eq:waveform}. These modes can be viewed as being sourced by the first-order ones, with their oscillation frequencies, damping times, and amplitudes expected to adhere to the relationships outlined below \cite{Ioka2007, Nakano2007, London2014}:
\begin{align}
	\label{eq:srm}
	\omega_{(l_1,m_1,n_1)(l_2,m_2,n_2)} &= \omega_{l_1,m_1,n_1} + \omega_{l_2,m_2,n_2}, \notag \\
	\tau_{(l_1,m_1,n_1)(l_2,m_2,n_2)}^{-1} &= \tau_{l_1,m_1,n_1}^{-1} + \tau_{l_2,m_2,n_2}^{-1}, \notag \\
	A_{(l_1,m_1,n_1)(l_2,m_2,n_2)} &\propto A_{l_1,m_1,n_1} A_{l_2,m_2,n_2}.
\end{align}
The ratio factor is defined as:
\begin{align}
	\mu_{(l_1,m_1,n_1)(l_2,m_2,n_2)} \equiv \frac{A_{(l_1,m_1,n_1)(l_2,m_2,n_2)}}{A_{l_1,m_1,n_1} A_{l_2,m_2,n_2}}.
\end{align}
This ratio factor is determined solely by the properties of the final black hole. Recently, Mitman et al. \cite{Mitman:2022qdl} and Cheung et al. \cite{Cheung:2022rbm} have verified the relationships in Eq. \eqref{eq:srm} using numerical relativity simulation results for head-on mergers and quasi-circular mergers with non-spinning progenitors, both obtaining similar values for the ratio factors of the (2,2,0)$\times$(2,2,0) and (2,2,0)$\times$(3,3,0) modes. In this paper, we adopt the ratio factors suggested by \cite{Cheung:2022rbm}:
\begin{eqnarray}
	\mu_{(2,2,0)(2,2,0)} = 0.1637, \notag \\
	\mu_{(2,2,0)(3,3,0)} = 0.4735.
\end{eqnarray}
Following the choices in both references, the spin-weighted spheroidal harmonics for the second-order QNMs used in this paper are:
\begin{align}
	\label{eq:srmswsh}
	_{-2}S_{(l_1,m_1,n_1)(l_2,m_2,n_2)}=_{-2}S_{l_1+l_2,m_1+m_2,n_1+n_2}
\end{align}

\subsection{Statistical method}
\label{sec:sta}
The main target of the paper is to investigate the spectroscopy of massive black hole with space-based \ac{GW} detectors, which will be represented by the detectability, resolvability, and waveform requirement of ringdown multiple \acp{QNM}. 

\paragraph{Detectability}
We use \ac{SNR} $\rho$ to to gauge the detectability of the multiple QNMs. In this study, we adhere to a detection threshold of $\rho=8$, aligning with the choices presented in \cite{Shi2019,Baibhav:2018rfk,Berti:2016lat}, and similar selections with varying values as seen in \cite{Ota:2019bzl,Divyajyoti:2021uty,Cabero:2019zyt}. The SNR for the ringdown signal is defined as::
\begin{align}
	\label{eq:snrrd}
	\rho_{Rd}\equiv\sqrt{(h_{Rd}|h_{Rd})},
\end{align}
where $h_{Rd}$ is the waveform of the ringdown stage, as delineated in Eq. \eqref{eq:waveform}. The SNR for any individual QNM $\rho_{l,m,n}$ is given by::
\begin{align}
	\label{eq:snrqnm}
	\rho_{l,m,n}\equiv\sqrt{(h_{l,m,n}|h_{l,m,n})}.
\end{align}
with $h_{{l,m,n}}$ representing the waveform considering only the $\{l,m,n\}$ mode. The inner product $(\dots|\dots)$ is defined as \cite{Finn:1992,Cutler:1994}
\begin{align}
	\label{rq:inner}
    (p(f)|q(f)) = 2\int_{f_\mathrm{low}}^{f_\mathrm{high}}
    \frac{p^*(f)q(f)+p(f)q^*(f)}{S_n(f)}df.
\end{align}
Where $p(f)$ and $q(f)$ are a pair of frequency-domain signals derived from their time-domain counterparts via Fourier transformation:
\begin{align}
	h(f)=\int^{+\infty}_{-\infty}h(t)e^{-2\pi ift}\mathrm{d}t,
\end{align}

To mitigate spurious power from the Fourier transformation, the frequency cutoffs $f_\mathrm{low}$ and $f_\mathrm{high}$ are set to half and twice the frequencies of the lowest and highest oscillation modes, respectively. $S_n(f)$ denotes the sky-averaged sensitivity of the detectors. For the TianQin detector, we adopt the following model \cite{Luo2016,Huang:2023prq,Ye:2023aeo,Bao:2019kgt,Zi:2021pdp,Huang:2024ylf,Sun:2022pvh,Lin:2023ccz,Xie:2022wkx,Shi:2022qno},
\begin{align}
	&S_n(f) = \frac{S_N(f)}{\overline{R}(f)},\notag\\
	&\overline{R}(f) \simeq  \frac{3}{10}
	\left[1+\left(\frac{2fL_0}{0.41c}\right)^2\right]^{-1},\notag\\
    &S_N(f) = \frac{1}{L^2}\left[\frac{4S_a}{(2\pi f)^4}
    \left(1+\frac{10^{-4}\rm{Hz}}{f}\right)+S_x\right].
\end{align}
Where $S_N(f)$ is the power spectrum density of detector noise, $\overline{R}$ is the sky averaged response-function, and $c$ is the speed of light, $L_0=\sqrt{3}\times10^8\mathrm{m}$ is the arm length of TianQin. $\sqrt{S_a}=1\times10^{-15}\mathrm{ms^{-2}Hz^{-1/2}}$ is the average residual acceleration on a test mass, and $\sqrt{S_x}=1\times10^{-12}\mathrm{mHz^{-1/2}}$ is the total displacement noise in a single link. For LISA,  we utilize the sensitivity curve provided in \cite{Robson2019}.

\paragraph{Resolvability}

The resolvability of any QNM should ensure that the target QNM parameters, specifically $\omega_{l,m,n}$ and $\tau_{l,m,n}$, can be discerned from other stronger modes. This criterion is based on the posterior distributions for these parameters satisfying the minimum separation requirement by the Rayleigh criterion. In essence, the resolvability criterion for the oscillation frequency and decay time of any 
$(l,m,n)$ mode is expressed as:
\begin{align}
	\label{eq:resolve}
	\delta\omega_{(l,m,n)}&<\min[\lvert \omega_{(l,m,n)}-\omega_{(l_1,m_1,n_1)}\rvert],\notag\\
	\delta\tau_{(l,m,n)}&<\min[\lvert \tau_{(l,m,n)}-\tau_{(l_2,m_2,n_2)}\rvert].
\end{align}
Where $(l_1,m_1,n_1)$ and $(l_2,m_2,n_2)$ are the modes stronger (as defined by \ac{SNR}) than $(l,m,n)$. $\delta\omega$ and $\delta\tau$ represent the parameter estimation accuracies for $\omega$ and $\tau$, respectively, which in this paper we compute via \ac{FIM}. 

For the case of large \ac{SNR} signals and Gaussian noise, the parameter estimation accuracies for $\theta_\alpha$ can be approximated by
\begin{align}
	\label{eq:pe}
	\delta\theta_\alpha\simeq\sqrt{(\Gamma^{-1})_{\alpha\alpha}},
\end{align}
where $\Gamma^{-1}$ is the inverse of \ac{FIM}, defined as:
\begin{align}
	\label{eq:fim}
	\Gamma_{\alpha\beta}\equiv\big(\frac{\partial h}{\partial\theta^\alpha}|\frac{\partial h}{\partial\theta^\beta}\big).
\end{align}
In this paper, we handle a 26-dimensional parameter space:
\begin{align}
	\label{eq:ps}
	\theta=\{M_z,D_L,\iota,\eta,\omega_{(l,m,n)},\tau_{(l,m,n)}\}.
\end{align}

\paragraph{Waveform modeling requirement}

We use mismatch to assess the waveform requirement for multiple QNMs. The mismatch between two waveforms  $h_1$ and $h_2$ is defined as
\begin{equation}
	\label{eq:mis}
\mathcal{M}=1-\frac{(h_1|h_2)}{\sqrt{(h_1|h_1)(h_2|h_2)}}
\end{equation}
where 
$h_1$ is the waveform obtained by summing all 11 QNMs, considered as the "true" ringdown waveform, and 
$h_2$ is the waveform obtained by summing partial QNMs, considered as the imperfect waveform used in data analysis.

A mismodeled waveform can introduce systematic errors that impact parameter estimation. It is crucial to minimize these systematic errors to be less significant than the statistical errors caused by instrumental noise. In this paper, we introduce a threshold $\mathcal{T}$ for mismatch $\mathcal{M}$, asserting that systematic errors are smaller than statistical errors when $\mathcal{M} < \mathcal{T}$. For the case of a large SNR signal, $\mathcal{T}$ is given by \cite{Chatziioannou2017}
\begin{equation}
\mathcal{T} = \frac{D}{2\rho_{Rd}^2},
\label{eq:threshold}
\end{equation}
where $D$ is the number of parameters whose measurability is affected by the model inaccuracy. For our model with 2 intrinsic parameters, $D=2$.

\section{Results}
\label{sec:result}
This section presents the findings on the detectability, resolvability, and waveform requirements for QNMs of the ringdown phase. Additionally, we incorporate results derived from astrophysical population models for massive black holes. For the sake of simplicity, we set the azimuthal angle $\phi=0$.

\subsection{Detectability}
\label{sec:de}
We commence our analysis by examining the parameter dependence of the detectability of TianQin and LISA on these 11 QNMs. Based on the definition of the SNR in Eq. \eqref{eq:snrqnm} and the ringdown waveform in Eq. \eqref{eq:waveform}, it is evident that the results are significantly influenced by the remnant black hole mass 
$M_z$, luminosity distance $D_L$, mass ratio $q$, and inclination angle $\iota$. However, the effect of $D_L$ on the results is solely due to its inverse proportionality to the amplitude, implying that the SNRs are inversely proportional to $D_L$ for all modes, a parameter we will not discuss further.

\begin{figure*}[htb]
		\subfigure{
			\includegraphics[scale=0.7]{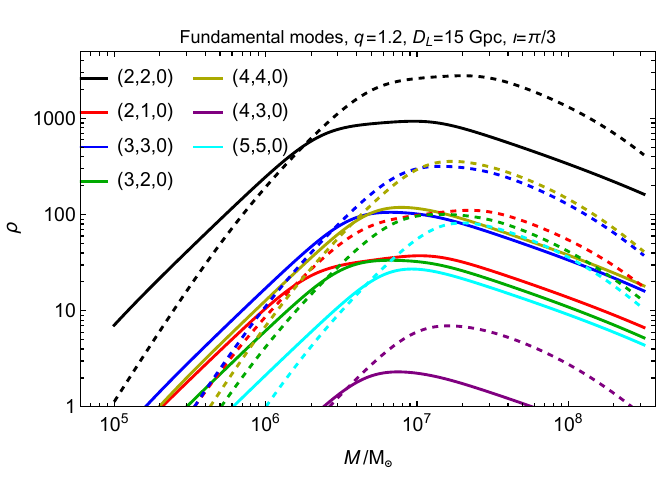}}
		\subfigure{
			\includegraphics[scale=0.7]{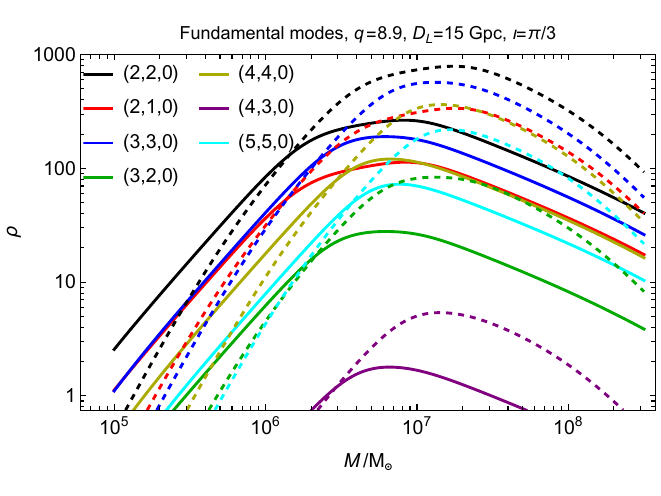}}
		\subfigure{
				\includegraphics[scale=0.7]{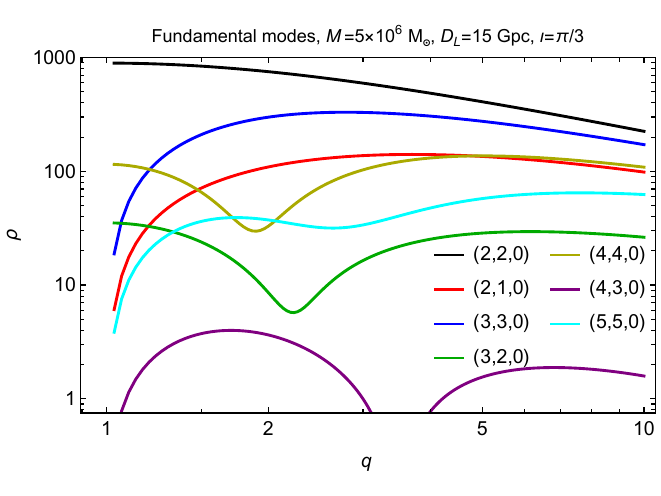}}
		\subfigure{
				\includegraphics[scale=0.7]{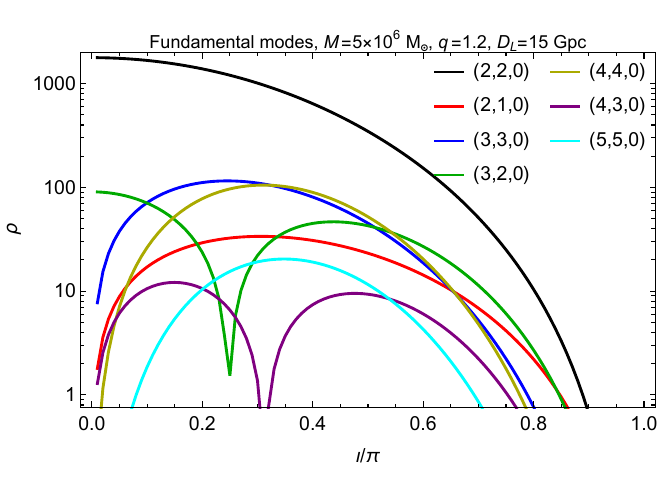}}
	\caption{\label{fig:snr_FM}
		Parametric dependence of the SNRs for fundamental modes on mass $M_Z$, mass ratio $q$ and inclination angle $\iota$.}
\end{figure*}

\begin{figure*}[htb]
	\subfigure{
		\includegraphics[scale=0.75]{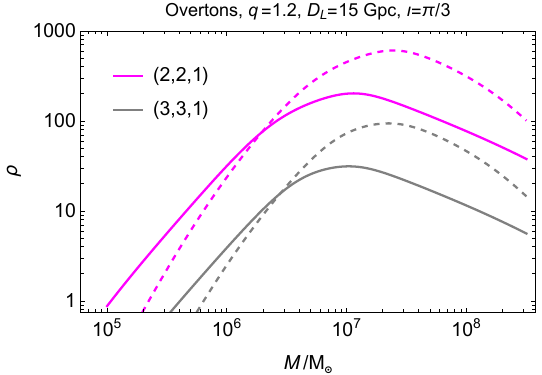}}
	\subfigure{
		\includegraphics[scale=0.75]{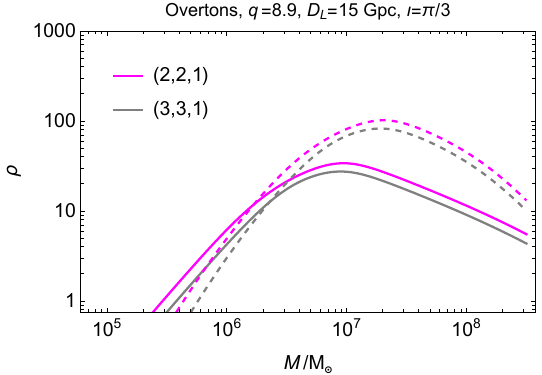}}
	\subfigure{
		\includegraphics[scale=0.75]{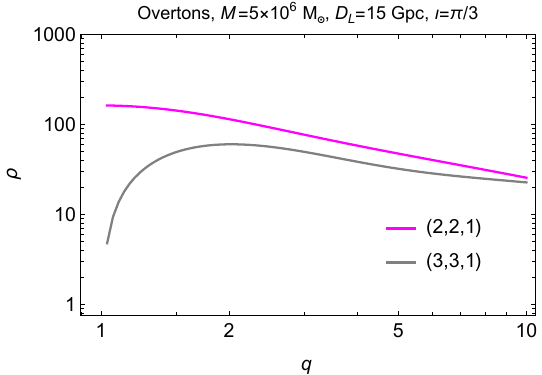}}
	\subfigure{
		\includegraphics[scale=0.75]{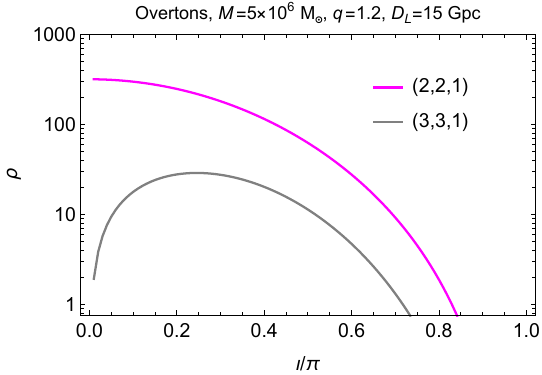}}
	\caption{\label{fig:snr_OT}
		Similar with Fig.\ref{fig:snr_FM} but for overtones.}
\end{figure*}

\begin{figure*}
	\subfigure{
		\includegraphics[scale=0.75]{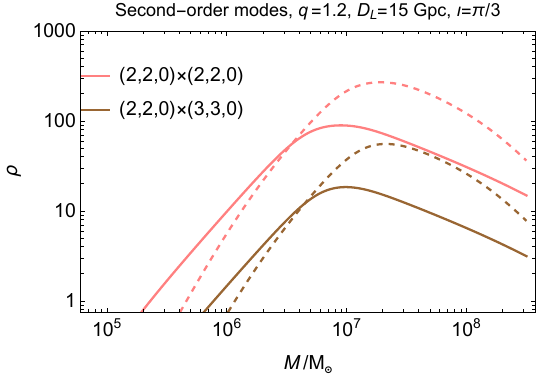}}
	\subfigure{
		\includegraphics[scale=0.75]{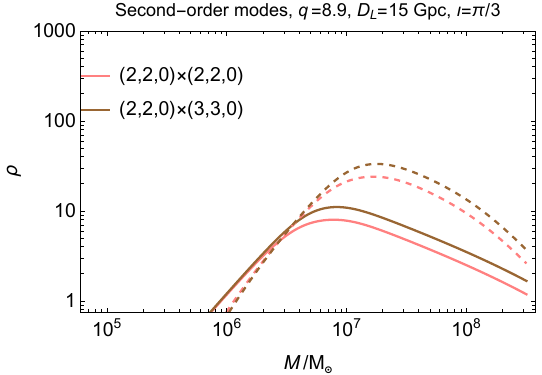}}
	\subfigure{
		\includegraphics[scale=0.75]{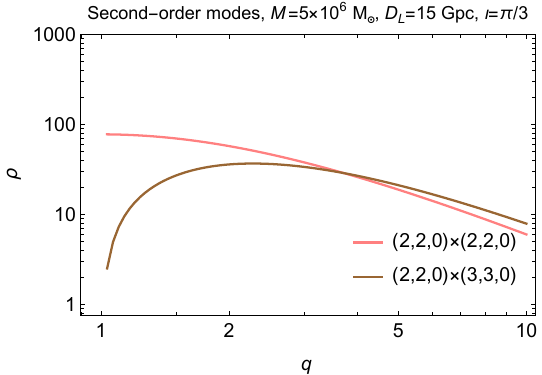}}
	\subfigure{
		\includegraphics[scale=0.75]{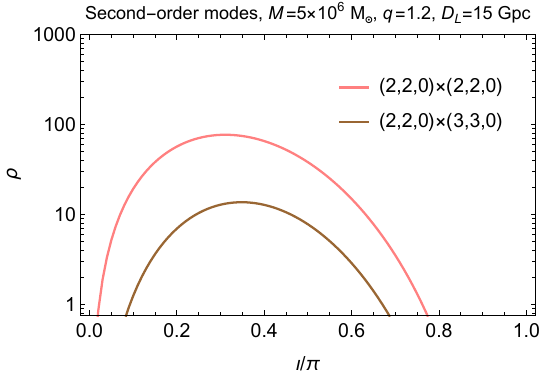}}
	\caption{\label{fig:snr_SM}
		Similar with Fig.\ref{fig:snr_FM} but for second-order modes.}
\end{figure*}

Our primary results are categorized into three groups: fundamental modes, overtones, and second-order modes, corresponding to Figures \ref{fig:snr_FM}, \ref{fig:snr_OT}, and \ref{fig:snr_SM}, respectively. For each category, we discuss the mass $M_z$ dependence under two scenarios: large mass ratio (exemplified by the mass ratio of GW190814, the source with the largest mass ratio in the current available GW data) and small mass ratio (exemplified by the mass ratio of GW150914). The top two plots in each figure illustrate these scenarios, with other parameters set to  $D_L=15\rm\  Gpc$ and $\iota=\pi/3$. It is evident that the impact of the total mass on SNRs is quite substantial. This effect stems from two aspects: one is the contribution of amplitudes,
\begin{align}
	\label{eq:snr_m1}
	\rho_{(l,m,n)}\propto \frac{M_z}{D_L}A_{(l,m,n)} S_{(l,m,n)} \propto M_z, 
\end{align}
which improves monotonically with $M_z$. The other effect is the influence of mass on multiple QNM frequencies and the performance of detectors at different frequencies. For space-based GW detectors like TianQin and LISA, the sky-averaged sensitivities tend to improve and then deteriorate with frequency. The results indicate that for all modes, the SNRs increase linearly with mass initially, reach a maximum value in the range of about $[10^6\rm M_\odot,10^7 M_\odot]$, and then decrease linearly. These figures also suggest that the higher the frequency of the QNM, the greater the total mass of the source corresponding to the peak SNR.

Given the nearly an order of magnitude difference in the sensitivity bands of these detectors, we highlight the strengths and weaknesses of TianQin and LISA in terms of their detectability of QNMs, represented by solid and dashed lines in the top two plots of Figures \ref{fig:snr_FM}, \ref{fig:snr_OT}, and \ref{fig:snr_SM}, respectively. Due to the differences in oscillation frequency of different modes, there will be a slight variation in the dominant region between TianQin and LISA. Overall, TianQin has an advantage in systems below $10^6 \rm M_\odot$, while LISA excels in systems above $10^7 \rm M_\odot$.
	

The dependences of \acp{SNR} on the mass ratio $q$ are primarily determined by the fitting formulas for the amplitude of each mode. This dependence is clearly demonstrated in the bottom left plots of Figures \ref{fig:snr_FM}, \ref{fig:snr_OT}, and \ref{fig:snr_SM}. Overall, the impact of the mass ratio on SNRs is not excessively severe, varying within an order of magnitude. The trends can be broadly categorized into three distinct types. Firstly, there are modes such as (2,2,0), (2,2,1), and (2,2,0)$\times$(2,2,0), where the SNR monotonically decreases with increasing mass ratio. Secondly, modes like (3,3,0), (2,1,0), (5,5,0), (3,3,1), and (2,2,0)$\times$(3,3,0) are difficult to detect in the equal-mass scenario. However, as the mass ratio approaches the 2-3 interval, they reach the optimal detection parameter space and then monotonically decrease. Finally, there are modes such as 320, 440, and 430, where the trend is relatively complex, often exhibiting a minimum in the mass ratio range of 2-3.

The dependences of \acp{SNR} on the inclination angle $\iota$ arise from the \acp{SWSH} associated with each mode. This dependence is illustrated in the bottom right plots of Figures \ref{fig:snr_FM}, \ref{fig:snr_OT}, and \ref{fig:snr_SM}. The plots indicate that an inclination angle greater than approximately 0.8 severely limits the detection capabilities of all modes. Additionally, all modes except (2,2,0), (2,2,1), and (4,3,0), the detection capabilities are also hindered when the angular momentum of the system is aligned with the line-of-sight direction. Overall, within the angle range of 0.1$\pi$ to 0.6$\pi$, all modes except (3,2,0) and (4,3,0) exhibit considerable detection capability. Notably, modes (3,2,0) and (4,3,0) display a minimum in detection capability around 0.25$\pi$ and 0.36$\pi$, respectively.

\subsection{Resolvability}
\label{sec:resolve}
The previous subsection focused on the detectabilities of detectors for multiple QNMs via the SNR, essentially comparing the strength of QNMs with detector noise. This subsection presents the resolvability results of each mode based on the criterion in Eq. \eqref{eq:resolve}.

According to the previous definition, the resolvability of a given mode depends on whether the posterior distribution of its oscillation frequency or decay time overlaps with those of other stronger modes. The computational procedure for determining the resolvability of a given mode 
$(l,m,n)$ in an event A is as follows: First, calculate the SNR of each mode for event A, sort the modes by SNR, and identify all modes stronger than $(l,m,n)$. Second, calculate the estimation accuracy of these modes through the Fisher information matrix. The final step is to judge whether the mode is resolvable based on Criterion Eq. \eqref{eq:resolve}.

\begin{figure*}
	\subfigure{
		\includegraphics[scale=0.75]{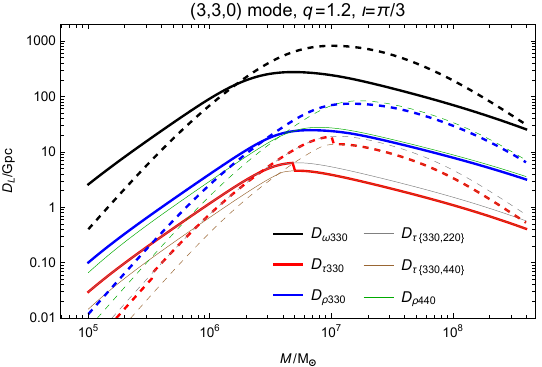}}
	\subfigure{
		\includegraphics[scale=0.75]{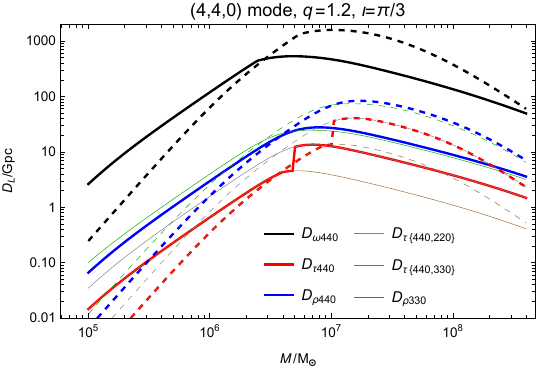}}
	\subfigure{
		\includegraphics[scale=0.75]{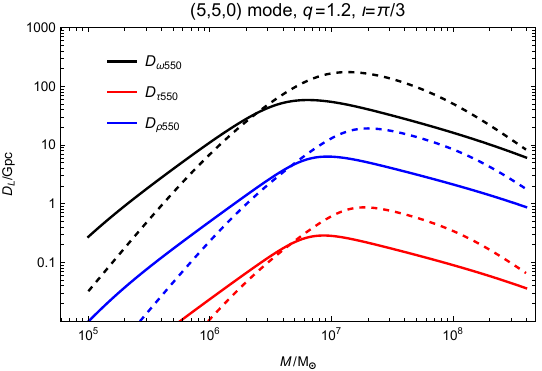}}
	\subfigure{
		\includegraphics[scale=0.75]{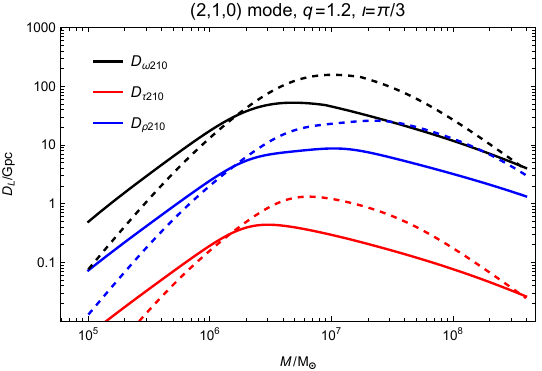}}
	\subfigure{
			\includegraphics[scale=0.75]{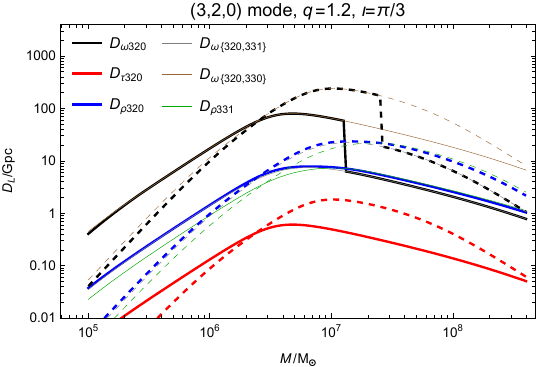}}
	\subfigure{
			\includegraphics[scale=0.75]{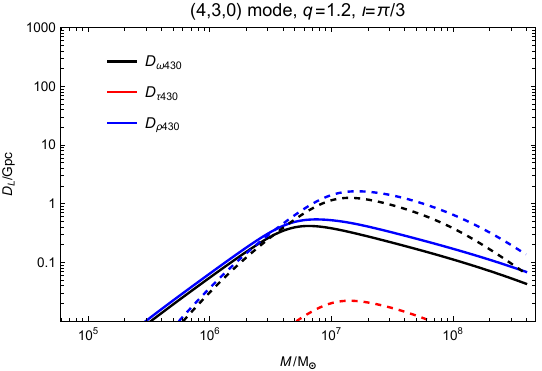}}
	\caption{\label{fig:resolve_FM}
		The dependence of the resolvabilities for fundamental modes on mass $M$.}
\end{figure*}

\begin{figure*}
	\subfigure{
		\includegraphics[scale=0.75]{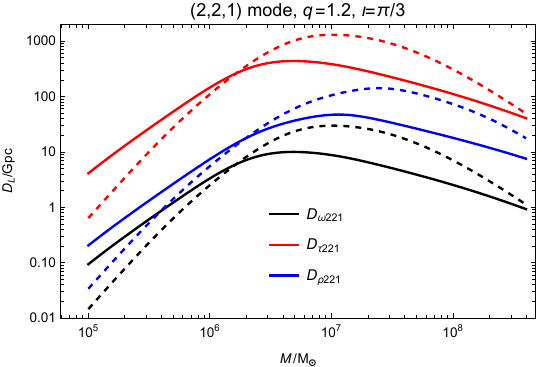}}
	\subfigure{
		\includegraphics[scale=0.75]{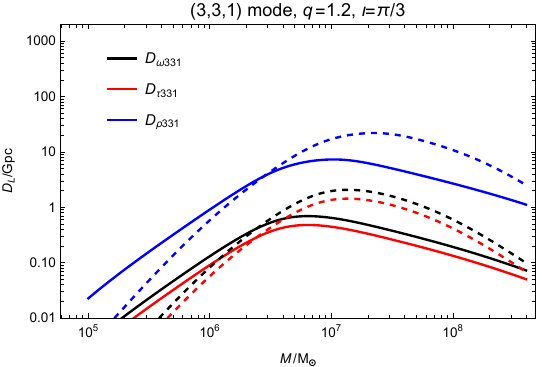}}
	\caption{\label{fig:resolve_OT}
		Similar with Fig.\ref{fig:resolve_FM} but for overtones.}
\end{figure*}

\begin{figure*}
	\subfigure{
		\includegraphics[scale=0.75]{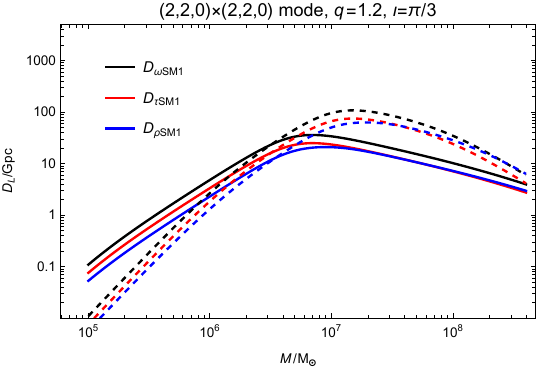}}
	\subfigure{
		\includegraphics[scale=0.75]{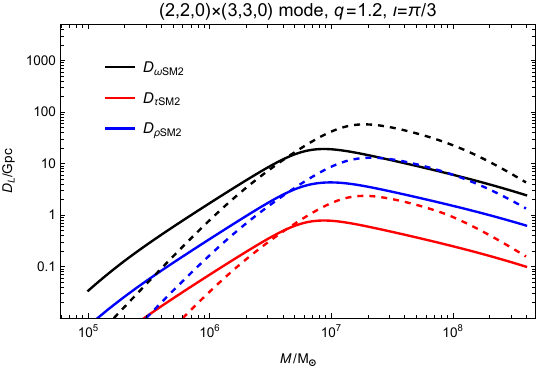}}
	\caption{\label{fig:resolve_SM}
		Similar with Fig.\ref{fig:resolve_FM} but for second order modes.}
\end{figure*}

We begin our analysis by examining the relationship between the resolvability of each mode and the source parameters. In this paper, we focus on the significant influence of the black hole mass $M_z$. The results are presented in terms of horizon distances, with key findings graphically represented in Figures \ref{fig:resolve_FM}, \ref{fig:resolve_OT}, and \ref{fig:resolve_SM}, corresponding to fundamental modes, overtones, and second-order modes, respectively. Other parameters are set to $q=1.2$ (GW150914-like), $D_L=15\rm Gpc$ and $\iota=\pi/3$. 

For each mode, we plot the anticipated resolvability of TianQin (solid line) and LISA (dashed line) on the oscillatory frequency (black, labeled as$D_{\omega lmn}$) and the decay time (red, labeled as $D_{\tau lmn}$). Additionally, we include the horizon distance at which each mode is detectable with an SNR of $\rho=8$ (blue, labeled as $D_{\rho lmn}$.

Notably, plots for modes (3,3,0), (3,2,0), and (4,4,0) exhibit discontinuities due to sorting modes by SNR during computation. For instance, mode (3,3,0) ranks second only to the (2,2,0) mode until the black hole mass approaches $5\times10^6 M_\odot$. Beyond this mass threshold, the (4,4,0) mode surpasses (3,3,0) in strength, necessitating the inclusion of (4,4,0) mode parameters in the horizon distance calculation for (3,3,0). The transition occurs because the decay time of mode (4,4,0) is more closely aligned with mode (3,3,0) than mode (2,2,0) ($|\tau_{330}-\tau_{220}|>|\tau_{330}-\tau_{440}|$), and the lower precision in estimating $\tau_{440}$ compared to (3,3,0) and (2,2,0) ($\delta\tau_{440}>\max[ \delta\tau_{220}, \delta\tau_{330}]$), resulting in a jump at approximately 
$5\times10^6 M_\odot$. The graph further illustrates alternative scenarios: one where (4,4,0) is consistently weaker than (3,3,0) (thin gray line, labeled 
$D_{\tau\{330,220\}}$), another where it’s consistently stronger (thin brown line, $D_{\tau\{330,440\}}$).

From Figure. \ref{fig:resolve_FM}, we find that for most fundamental modes, except for the (4,3,0) mode, the oscillation frequency resolvable horizon distance is almost an order of magnitude better than the detection horizon distance with a SNR of 8, while the decay time resolvable horizon distance is an order of magnitude worse. For the 430 mode, the oscillation frequency is very close to that of the $(2,2,0)\times(2,2,0)$ mode, and the $(2,2,0)\times(2,2,0)$ mode tends to be stronger than the 430 mode, leading to an order of magnitude depressions in the oscillation frequency resolvable horizon distance. 

For overtones, i.e. Figure \ref{fig:resolve_OT}, the oscillation frequencies are very close to the corresponding fundamental modes, and the accuracy of the frequency measurements of these modes are much worse than that of their fundamental modes, leading to a significant depression (1-2 orders of magnitude) of the resolvable horizon distance of the oscillation frequency of these modes. For (2,2,1) modes, modes that are stronger than this mode mainly include (2,2,0), (3,3,0), and (2,1,0) modes, which are very different in terms of the decay time, leading to a significant increase in the resolvable horizon distance of decay time. 

In the context of second-order modes, as shown in Figure \ref{fig:resolve_SM}, the resolvable horizon distance for oscillation frequency and damping time of (2,2,0)$\times$(2,2,0) mode are very close to the detectable horizon distance. Notably, the frequency of this mode is nearly identical to the 440 mode, but under the such parameter selection, it is weaker, resulting in a shortened resolvable horizon distance for oscillation frequency. Conversely, the decay time of the (2,2,0)$\times$(2,2,0) mode stands apart from stronger modes, contributing to an expansion of the resolvable horizon distance.

Upon analyzing the three diagrams presented, it becomes evident that TianQin surpasses LISA in terms of detection capability, oscillation frequency resolvability, and decay time resolvability within the low-mass range. Conversely, LISA exhibits superiority over TianQin in the large-mass range. It can also be noticed from these figures that, aside from the dominant (2,2,0) mode, only the (3,3,0), (4,4,0), and (2,2,1) modes exhibit notable detectability and resolvability on both $\omega$ and $\tau$, whereas other modes struggle to concurrently resolve both two parameters. Specifically, the (4,3,0) and (3,3,1) modes both exhibit relatively small resolvable horizon distances for both oscillation frequencies and decay times, indicating that they will be difficult to resolve from other stronger modes.

Massive black hole population models anticipate that TianQin and LISA could detect from tens to hundreds of massive black hole mergers throughout their mission periods \cite{Wang:2019ryf,Feng:2019wgq}. However, the exact number and characteristics of these mergers detectable by these two detectors are heavily influenced by the underlying astrophysical models. In this study, we have delved into the detectability and resolvability of QNMs for TianQin and LISA, considering three distinct astrophysical scenarios: "popIII," "Q3\_d", and "Q3\_nod". These scenarios correspond to a light seed model \cite{Madau:2001sc} and two heavy seed models \cite{Bromm:2002hb,Begelman:2006db,Lodato:2006hw}, with and without time delay between the mergers of massive black holes and their host galaxies, respectively.

To evaluate these scenarios, we have generated 1000 simulated catalogues of observable events for each of the three models. Each catalogue encapsulates all potential events that could be detected by TianQin and LISA, assuming a detection threshold of 8. By averaging these 1000 mock catalogues for each detector, we have derived the expected detection and resolution number, with the comprehensive results presented in Table \ref{tablemr}.

Our analysis reveals that, with the exception of the (3,3,1) and (4,3,0) modes, all other QNM modes, including two second-order modes, exhibit encouraging detectability and resolvability prospects. This conclusion resonates with the outcomes of our parametric-dependence analysis, further strengthening the validity and significance of our findings.


\begin{table*}[!htbp]
	\label{tablemr}
	\renewcommand{\arraystretch}{1.7}
	\centering
	\begin{tabular}{|p{1.4cm}|p{1.8cm}|p{0.9cm}<{\centering}|p{0.9cm}<{\centering}|p{0.9cm}<{\centering}|p{0.9cm}<{\centering}|p{0.9cm}<{\centering}|p{0.9cm}<{\centering}|p{0.9cm}<{\centering}|p{0.9cm}<{\centering}|p{0.9cm}<{\centering}|p{2cm}<{\centering}|p{2cm}<{\centering}|}
		\hline
		\multirow{2}{*}{\makecell[l]{Type}}&\multirow{2}{*}{\makecell[l]{Cases}}&\multicolumn{11}{c|}{Quasi normal mode}\\
		\cline{3-13}
		&&(2,2,0)&(2,2,1)&(2,1,0)&(3,3,0)&(3,3,1)&(3,2,0)&(4,4,0)&(4,3,0)&(5,5,0)&(2,2,0)$\times$(2,2,0)&(2,2,0)$\times$(3,3,0)\\
		\hline
		\multirow{6}{*}{\makecell[l]{Detection\\ \\Number}} & PIII\_TQ & 11.6 & 4.8 & 7.2 & 7.7 & 2.5 & 2.9 & 5.0 & 1.0 & 2.8 & 2.2 & 1.2\\ \cline{2-13}
		&Q3d\_TQ & 13.7 & 8.3 & 8.4 & 9.8 & 3.4 & 3.4 & 6.7 & 0.8 & 2.5 & 4.5 & 1.4\\\cline{2-13}
		& Q3nod\_TQ & 168.1 & 59.0 & 78.5 & 110.2 & 16.6 & 16.9 & 46.8 & 4.5 &
		16.9 & 15.0 & 5.6\\\cline{2-13}
		&PIII\_LISA& 20.7 & 11.2 & 15.3& 14.8& 7.4 & 9.3 & 10.9 & 4.6 & 7.3 & 6.1 & 4.0 \\\cline{2-13}
		&Q3d\_LISA & 20.1& 11.5& 17.0 & 16.3& 7.4& 9.9 & 11.8& 4.1& 7.4& 5.2& 3.0 \\\cline{2-13}
		&Q3nod\_LISA & 296.4 & 136.6 & 204.6 & 207.6 & 62.2 & 76.6 & 113.8 & 26.9& 53.4 & 51.1 & 24.1 \\\cline{1-13}
		\multirow{6}{*}{\makecell[l]{Resolution\\ \\Number}} & PIII\_TQ & 11.6& 4.5 & 3.9 & 6.9 & 0.1 & 1.2 & 4.6 & 0.1 & 2.5 & 1.0 &	0.8 \\\cline{2-13}
		& Q3d\_TQ & 13.7 & 8.2 &, 6.2 & 9.5 & 0.1 & 2.4 & 6.5 & 0.1 & 2.2 & 2.4 & 1.0 \\\cline{2-13}
		& Q3nod\_TQ & 168.1 & 57.1 &38.2& 106.1& 0.3 & 6.1 & 44.7 & 0.5 & 14.3 & 6.8 &4.1  \\\cline{2-13}
		& PIII\_LISA & 20.7 & 9.8 & 9.7& 13.2 & 0.3 & 4.5 & 9.6 & 0.5 & 6.3 & 3.1 &	2.8   \\\cline{2-13}
		& Q3d\_LISA  &  20.1 & 10.6 & 12.5 & 14.5 & 0.2 & 4.2 & 10.4 & 0.4 & 6.4 & 2.6 & 2.0   \\\cline{2-13}
		& Q3nod\_LISA &  296.4 & 129.2 & 143.5 & 195.3 & 1.8 & 32.7 & 106.5 & 3.0 & 45.7&	24.2 & 17.0   \\\cline{2-13}
		\hline
	\end{tabular}
	\caption{Main results about detection number and resolve number for different scenarios are list in this table.}
\end{table*}

\subsection{Modeling requirement}
\label{sec:modeling}
The above results suggest that TianQin and LISA may detect and resolve multiple \acp{QNM}, but resolving the (3,3,1) and (4,3,0) modes, particularly the (4,3,0) mode with an SNR unlikely to surpass 8, poses challenges. This subsection delves further into the necessity of incorporating the (3,3,1) and (4,3,0) modes into the ringdown waveform templates for the high-precision modeling requirements of space-based \ac{GW} detector.

Utilizing the mismatch (defined in Eq. \eqref{eq:mis}) and its associated threshold (specified in Eq.\eqref{eq:threshold}), we conducted separate analyses to evaluate the significance of including the (4,3,0) and (3,3,1) modes in the modeling of ringdown waveforms. In this context, the physical waveform, denoted as $h_1$ in Eq. \eqref{eq:mis}, serves as the comprehensive or complete waveform and is constructed as the linear superposition of all 11 modes discussed in this study, i.e., $h_1\equiv h_{Rd}$. On the other hand, the template waveform, labeled $h_2$, is constructed by excluding either the (3,3,1) or (4,3,0) mode, thereby representing the linear combination of the remaining 10 modes, i.e., $h_{2,lmn}\equiv h_{Rd}-h_{lmn}$.

\begin{figure*}[htb]
	\subfigure{
		\includegraphics[scale=0.75]{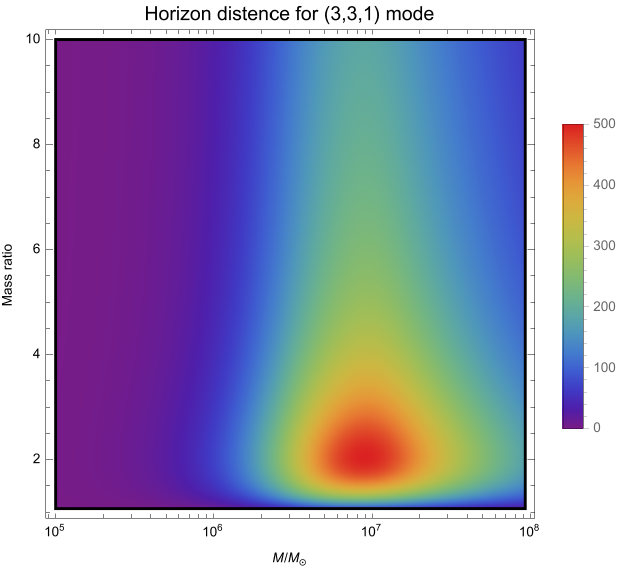}}
	\subfigure{
		\includegraphics[scale=0.75]{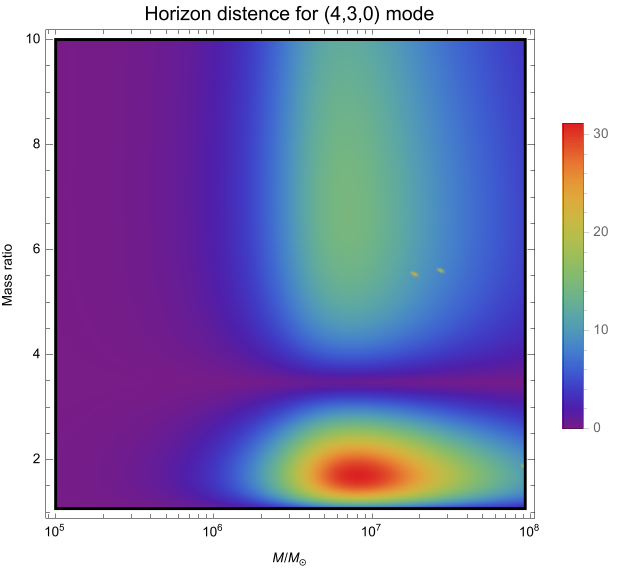}}
	\caption{The dependence of the waveform modeling requirement for (3,3,1) and (4,3,0) modes on mass $M$ and mass ratio $q$.}
	\label{fig:mis}
\end{figure*}

We conducted a detailed analysis of how the modeling requirements vary with the black hole mass and the initial mass ratio of the binary black hole system by taking TianQin as example. The significance of including or excluding a particular mode in the modeling is quantified by the luminosity distance at which the mismatch caused by the absence of this mode reaches threshold. The key results are presented in Figure. \ref{fig:mis}. One can conclude from this figure, while the (3,3,1) mode poses a challenge in being resolved from other more dominant modes, its absence during the modeling process is unacceptable. Similarly, for the 430 mode, despite its rarity in exceeding \ac{SNR} of 8 and its considerable difficulty in being resolved from the ringdown signals, the systematic errors introduced by its omission cannot be overlooked. This is especially crucial for systems massive than $10^6 \rm M_\odot$, and in the case of massive binary black hole systems located at a distance of 10 Gpc apart from the detector, the inclusion of this mode in the waveform templates is of paramount importance and must be carefully considered.

\section{Conclusion}
\label{sec:conclusion}
In this paper, we have conducted a thorough analysis of the detectabilities and resolvabilities of space-based GW detectors, with a focus on TianQin and LISA, for multiple QNMs during the ringdown phase of massive binary black hole mergers. Our study, constrained by the current amplitude fitting results of QNMs as detailed in \cite{London2014} and \cite{Cheung:2022rbm}, has centered on 11 specific QNMs. By assessing the SNR and parameter estimation accuracy, we have evaluated the capabilities of these detectors in detecting and resolving these modes. Additionally, we have discussed the dependence of these capabilities on source parameters, and we estimate the expected detection and resolution numbers for TianQin and LISA for by utilizing three astrophysical population models of massive black holes.

Our results demonstrate that, aside from the (3,3,1) and (4,3,0) modes, both detectors show considerable potential in detecting and resolving the remaining nine modes, including the one overtone mode (2,2,1) and two nonlinear second-order modes (2,2,0)$\times$(2,2,0) and (2,2,0)$\times$(3,3,0). The (3,3,1) mode, despite its relatively favorable SNR, faces significant challenges due to its close frequency proximity to the (3,3,0) mode, complicating its resolution from the ringdown signal. The (4,3,0) mode, characterized by its weak signal strength, seldom achieves an SNR above 8 in TianQin and LISA, further complicating its resolution from stronger modes. However, our mismatch analysis underscores the critical importance of including these modes in the construction of ringdown waveform templates for space-based GW detectors. Even the weakest (4,3,0) mode, if omitted, can introduce substantial systematic errors into the ringdown signals, particularly for systems with a total mass greater than $M>10^6 \rm M_\odot$

While our study provides valuable insights, it is not without limitations. Firstly, our analysis is limited to only 11 modes, and there is a clear need to introduce and discuss a broader range of QNMs, including fundamental modes, overtones, and second-order modes. Secondly, the amplitude fitting utilized in this paper is applicable to initial no-spin binary black hole systems and serves as a basis for preliminary analysis. However, more rigorous studies necessitate the use of amplitude results derived from more realistic scenarios. Lastly, the statistical methods employed in this paper for analyzing detectability and resolvability are not directly applicable to practical data processing situations. The development of appropriate data-processing tools, ideally based on Bayesian analysis, is essential to bridge the gap between theoretical analysis and real-world applications. And the SNR threshold used in the article is simply and roughly set to 8. As highlighted by Sun et al. \cite{Sun:2024nut}, the threshold for detecting higher-order effects predicted by general relativity may not require such a high value. Further investigation into determining a more precise threshold is warranted to refine our analysis

\begin{acknowledgements}
	The authors thank Yi-Ming Hu and Jian-dong Zhang for useful discussions. This work is supported by the Guangdong	Major Project of Basic and Applied Basic Research (Grant No. 2019B030302001)	and the National Science Foundation of China (Grant No. 12261131504).
\end{acknowledgements}

\bibliography{reference}% Produces the bibliography via BibTeX.

\end{document}